\documentclass[sigconf,authorversion,nonacm]{acmart}
\AtBeginDocument{%
  \providecommand\BibTeX{{%
    \normalfont B\kern-0.5em{\scshape i\kern-0.25em b}\kern-0.8em\TeX}}}

\acmConference[Preprint]{}{September}{2024}
\acmBooktitle{Preprint} 

\usepackage[polutonikogreek,english]{babel}
\usepackage[or]{teubner}
\usepackage{rotating}
\usepackage{graphicx}
\makeatletter
\def\Hline{
  \noalign{\ifnum0=`}\fi\hrule \@height 4.\arrayrulewidth \futurelet
   \reserved@a\@xhline}
\makeatother

\begin{document}

\title[AI as Extraherics]{AI as Extraherics: Fostering Higher-order Thinking Skills in Human-AI Interaction}


\author{Koji Yatani}
\affiliation{%
  \institution{The University of Tokyo}
  \streetaddress{7-3-1 Hongo}
  \city{Bunkyo-ku}
  \state{Tokyo}
  \country{Japan}
  \postcode{113-8656}
}
\email{koji@iis-lab.org}

\author{Zefan Sramek}
\affiliation{%
  \institution{The University of Tokyo}
  \streetaddress{7-3-1 Hongo}
  \city{Bunkyo-ku}
  \state{Tokyo}
  \country{Japan}
  \postcode{113-8656}
}
\email{zefanS@iis-lab.org}

\author{Chi-Lan Yang}
\affiliation{%
  \institution{The University of Tokyo}
  \streetaddress{7-3-1 Hongo}
  \city{Bunkyo-ku}
  \state{Tokyo}
  \country{Japan}
  \postcode{113-8656}
}
\email{chilan.yang@iii.u-tokyo.ac.jp }
\renewcommand{\shortauthors}{Yatani et al.}

\begin{abstract}
As artificial intelligence (AI) technologies, including generative AI, continue to evolve, concerns have arisen about over-reliance on AI, which may lead to human deskilling and diminished cognitive engagement.
Over-reliance on AI can also lead users to accept information given by AI without performing critical examinations, causing negative consequences, such as misleading users with hallucinated contents.
This paper introduces \textit{extraheric AI}, a human-AI interaction conceptual framework that fosters users' higher-order thinking skills, such as creativity, critical thinking, and problem-solving, during task completion.
Unlike existing human-AI interaction designs, which replace or augment human cognition, extraheric AI fosters cognitive engagement by posing questions or providing alternative perspectives to users, rather than direct answers.
We discuss interaction strategies, evaluation methods aligned with cognitive load theory and Bloom's taxonomy, and future research directions to ensure that human cognitive skills remain a crucial element in AI-integrated environments, promoting a balanced partnership between humans and AI.
\end{abstract}







\maketitle

\section{Introduction}
Recent advances in artificial intelligence (AI), including generative AI, have shown strong potential to support human tasks, reduce workloads, and augment capabilities, but concerns have arisen about the over-use or over-reliance on AI technology~\cite{buccinca2021trust, chen2023understanding}.
Such reliance on AI for cognitive tasks can lead to deskilling, where individuals lose opportunities for cognitive skill maintenance and development~\cite{KnowledgeWorkerCHI2024}.
Over-reliance on AI for information-seeking may also sway users toward particular viewpoints or opinions presented by AI, as seen in writing and design explorations~\cite{DraxlerTOCHI, WadinambiarachchiCHI2024}, and can further exacerbate issues of misinformation and disinformation when users blindly trust erroneous or hallucinated AI outputs.
This dependence may also diminish perceived ownership, sense of challenge, productivity, and accomplishment~\cite{KobiellaCHI2024}.
A fundamental issue underlying these negative consequences is the focus of current human-AI interaction research on supporting human tasks by replacing or augmenting human cognitive abilities.
Such AI design may enhance task efficiency but deprive users of opportunities for cognitive engagement and growth.
With generative AI becoming increasingly capable of outperforming humans in many tasks, users may be more likely to trust AI without skepticism.

To address these challenges, researchers have explored redesigning human-AI interactions to promote cognitive engagement.
Tankelevitch et al. discuss the potential of generative AI for expanding users' metacognitive capabilities~\cite{MetacognitionCHI2024}.
Danry et al. showed that asking questions about a user’s argument, rather than providing additional explanations, can stimulate users' critical thinking through \textit{``human-AI co-reasoning''}~\cite{10.1145/3544548.3580672}.
These discussions and projects suggest a strong potential for human-AI interaction research to stimulate users' creativity, critical thinking, and problem-solving skills among the seven core 21st-century skills outlined by van Laar et al.~\cite{vanLaar2017577}.

In this paper, we introduce a novel conceptual framework for human-AI interaction: \textit{extraheric AI}.
We define \textit{``extraherics''} as a mechanism that fosters users' higher-order thinking skills during the course of task completion.
\textit{Extraheric} is based on the Latin word ``extrah\={o}'' (to draw forth or pull out), and we use this term to suggest that AI can draw forth people's higher-order thinking skills and thus promote their cognitive potential.
Rather than replacing or augmenting human cognitive abilities, extraheric AI encourages users to engage in higher-order thinking during task completion.
For instance, in writing, extraheric AI might prompt users to reflect on specific content or visualize how others have approached similar topics, rather than directly performing revisions or replacement.
This process encourages users to examine, select, and synthesize information, creating implicit learning opportunities that foster higher-order thinking skills.

As users increasingly rely on intelligent systems, extraheric AI aligns with N. Katherine Hayles’ vision of a positive \textit{posthuman} future, where humans are not \textit{``hopelessly compromised''} by machines but instead engage in a collaborative cognitive environment with them~\cite{hayles1999posthuman}.
In this model, thinking is a shared process between human and nonhuman actors, ensuring that human higher-order thinking skills remain vital and are actively fostered~\cite{hayles1999posthuman}.

This paper provides an overview of extraheric AI, detailing its interaction strategies, evaluation methods, design considerations, and future research directions.
The contributions include:

\begin{itemize}
    \item Defining the novel human-AI interaction conceptual framework, \textit{extraheric AI}, and comparing it with other human-AI interaction designs using cognitive load theory,
    \item Identifying interaction strategies for extraheric AI based on recent HCI literature,
    \item Outlining evaluation methods for assessing users' cognitive, attitudinal, and behavioral change when using extraheric AI, and
    \item Exploring design considerations and proposing future research directions for extraheric AI.
\end{itemize}

\section{Related Work}

\subsection{Higher-Order Thinking Skills}

Lewis and Smith define higher-order thinking as a cognitive process that \textit{``occurs when a person takes new information and information stored in memory and interrelates and/or rearranges and extends this information to achieve a purpose or find possible answers in perplexing situations''}~\cite{lewis1993defining}.
This definition is purposefully broad, and, as the authors note, higher-order thinking can be used for a variety of tasks, including: \textit{``deciding what to believe; deciding what to do; creating a new idea, a new object, or an artistic expression; making a prediction; and solving a nonroutine problem''}~\cite{lewis1993defining}.
\textit{Critical thinking skills} and \textit{creative thinking skills} are both characteristic features of higher-order thinking skills~\cite{conklin2011higher}. 

Beyond the acquisition of knowledge and skills, the development of the ability to think critically has become a key educational goal in the last half-century~\cite{mcpeck1981critical}, and it is increasingly essential in the face of the development of advanced AI tools that are changing the ways we gather and interact with information.
Broadly speaking, \textit{critical thinking} describes \textit{``thinking that is purposeful, reasoned, and goal-directed---the kind of thinking involved in solving problems, formulating inferences, calculating likelihoods, and making decisions when the thinker is using skills that are thoughtful and effective for the particular context and type of thinking task''}~\cite{halpern1996thought}, as well as \textit{``the appropriate use of reflective skepticism within the problem area under consideration''}~\cite{mcpeck1981critical}.

Although often associated with artistic creation, \textit{creative thinking} is also broadly defined as thinking applicable to numerous domains.
Torrance defines creative thinking as \textit{``the process of sensing gaps or disturbing, missing elements; forming ideas or hypotheses concerning them; testing these hypotheses; and communicating the results, possibly modifying and retesting the hypotheses''}~\cite{torrance1964guiding}.
Similarly to critical thinking, creative thinking has been of increasing interest to educators over the last several decades.
Meador notes that \textit{``the ability to think creatively is essential in life for many reasons, including solving problems, producing meaningful and satisfying ideas and products, and developing works in art forms''}~\cite{meador1997creative}. 

Both critical and creative thinking are skills that can be developed by learning about and practicing reasoning, analysis, planning, and questioning~\cite{halpern1996thought,lai2011metacognition,michalko2010thinkertoys}.
Aside from such knowledge and practice, however, Halpern also emphasizes the importance of the \textit{attitudes} of a critical thinker, characterized by willingness to plan, flexibility, persistence, willingness to self-correct, being mindful, and consensus-seeking~\cite{halpern1996thought}.
Similarly, Michalko emphasizes the importance of cultivating a creative attitude, which includes the \textit{belief} that one is creative~\cite{michalko2010thinkertoys}.
Encouraging and developing such attitudes among users is also a key goal for extraheric AI because these attitudes allow users to transfer the skills they have developed to other task domains.

Metacognition is another concept related to and yet distinct from higher-order thinking.
Metacognition refers to ``thinking about thinking,'' being aware of one's cognitive processes and regulating them to achieve specific goals~\cite{lai2011metacognition}. 
Metacognitive skills include \textit{metacognitive knowledge} (awareness of one's own learning processes, strategies, and strengths or weaknesses)~\cite{Flavell1979} and \textit{metacognitive regulation} (organizing, monitoring, and assessing one's own learning activities)~\cite{Brown1978}.
Metacognitive skills may thus help different aspects of higher-order thinking; for example, when analyzing a complex problem, metacognitive skills help a person plan how to approach the problem, monitor their comprehension, and adjust their strategies as appropriate. 
In this sense, metacognition focuses on awareness and regulation of thinking processes, enabling effective use of higher-order thinking skills like analysis, evaluation, and creation.
Tankelevitch et al.~\cite{MetacognitionCHI2024} have recently explored the role of metacognition in the use of generative AI, arguing that generative AI places substantial metacognitive demands on the user.
They propose that generative AI systems can aid users by including metacognitive supports as well as reducing metacognitive demands through design, by \textit{``offload[ing] metacognitive processing from the user to the system''}~\cite{MetacognitionCHI2024}.
Extraheric AI proposes a different perspective on the division of cognitive labor between users and AI.
It aims at stimulating cognitive activities to foster higher-order thinking skills through the use of AI, instead of offloading such activities.

\subsection{Technology Support for Higher-Order Thinking Skills}
Higher-order thinking skills in general have received little attention from the HCI community to date, though some scholars from other research fields have examined the use of technology to foster higher-order thinking skills~\cite{hopson2001using}.
That being said, there exists a long history of scholarship on the ways that technology can be harnessed to promote \textit{critical thinking}, \textit{creative thinking}, and support educational goals in general.
Early research included discussions about the roles of computers in schools~\cite{roschelle2000changing}, and proposals, such as Jonassen's concept of \textit{Mindtools}~\cite{jonassen2000computers}.

HCI researchers have long been interested in creativity support tools~\cite{frich2019mapping}, and have developed numerous methods for evaluating their impacts on creative thinking~\cite{remy2020evaluating}.
Recent research in HCI and related fields has also explored various techniques for promoting critical thinking in a variety of application domains, including web search~\cite{yamamoto2018query}, online collaboration~\cite{sun2017critical}, educational exhibitions~\cite{10.1145/3544548.3581159}, online learning~\cite{inoue2023encouraging}, digital media literacy~\cite{perez2017media}, data sensing~\cite{lechelt2020coming}, engineering research~\cite{asiri2021assessing}, and misinformation mitigation~\cite{dingler2021workshop, boonprakong2023designing}. 
With the power and flexibility of AI, extraheric AI has the potential to accelerate this research direction and play a substantial role in the development and promotion of higher-order thinking skills.

Interactive Tutoring Systems (ITS) is a domain related to interaction and interface designs for higher-order thinking skill development\footnote{ITS is also known as Intelligent Computer-Aided Instructions (ICAI) in the domain of educational systems using AI, but the term of ITS has been adopted most widely~\cite{Nwana1990}.}.
ITS can foster students' higher-order thinking skills by providing personalized learning experiences, such as adaptive feedback and problem-solving exercises. 
ITS can also offer support for thinking activities by incorporating expert thinking models and metacognitive strategies~\cite{gama2004metacognition}.
While ITS is specifically designed to fit into educational contexts, extraheric AI can be integrated into users' existing non-educational activities.
Extraheric AI thus has the potential to enhance users' higher-order thinking skills in a broader set of application domains.
In an example of building abilities to counter mis/dis-information on the Internet, ITS can offer personalized teaching content based on students' preferences and behavior on online content. 
Extraheric AI can complement this by encouraging the execution of higher-order thinking skills when users read online content in a practical setting.

Effortful user interfaces are another interface design relevant to extraheric AI.
Effortful user interfaces purposefully infuse cognitive load into interfaces in order to encourage users' learning of computing systems~\cite{EffortfulCockburn2007}.
Effortful user interfaces typically introduce additional burdens during task completion, which serves as a motivator for users to learn.
For example, Grossman et al. revealed that disabling the activation of a command in a pull-down menu can lead to accelerated learning of keyboard shortcuts~\cite{GrossmanCHI2007}.
Similar concepts and merits of introducing purposeful workload to interface designs have been discussed within HCI~\cite{RichCHIEA2010}.
These interface designs take an approach to infuse deliberate intrinsic cognitive load to encourage users' learning and behavior. 
In a related way, extraheric AI instead aims to increase germane cognitive load to stimulate cognitive activities that can foster users' higher-order thinking skills, as described in Section~\ref{sec:definition}.

\section{Extraheric AI}
\label{sec:definition}

Human-AI interaction systems are typically designed to directly support human tasks, such as by taking on subtasks, accelerating processes, or reducing input effort.
In his book \textit{Human-Centered AI}, Ben Schneiderman offers the following categorization for tools serving human needs~\cite{ShneidermanHCAI}:

\begin{itemize}
 \item \textit{Orthotics}: Systems that enhance performance in specific tasks (e.g., auto-completion, FlashFill in Excel, and Copilot for coding).
 \item \textit{Prosthetics}: Systems that replace missing capabilities (e.g., real-time captioning, and visual information verbalization).
 \item \textit{Exoskeletons}: Systems that expand human capacities related to specific tasks (e.g., language translation, and information search assistants).
\end{itemize}

Many existing human-AI interaction systems fit one or more of these categories, although their classification can vary depending on context and user capabilities.
For example, a language translation application serves as prosthetics for users with no background in a language, but acts as exoskeleton for those with some proficiency.

Unlike these categories, extraheric AI focuses on fostering users' higher-order thinking skills through interaction.
Extraheric AI agents help users explore different information and perspectives while completing tasks, instead of providing direct support.
Extraheric AI thus introduces a new research direction for human-AI interaction: increasing germane cognitive load.
According to John Sweller’s cognitive load theory, there are three types of cognitive load: \textit{intrinsic}, \textit{extraneous}, and \textit{germane}~\cite{Sweller1988}.
Intrinsic load is the mental effort required by the inherent complexity of the material.
Extraneous load is the additional mental effort caused by how information is presented, which does not directly aid learning (e.g., poorly-designed interfaces).
Germane load is the mental effort involved in constructing and automating \textit{schemas} (cognitive frameworks that help organize information efficiently).
Increasing germane load is thus desirable because it fosters the development and retention of higher-order thinking skills.

Extraheric AI aims to increase germane cognitive load while the other AI types (orthotics, prosthetics, and exoskeleton) primarily focus on reducing intrinsic or extraneous cognitive load.
This also suggests that extraheric AI can coexist with these supporting AI designs.
For instance, a writing support AI system might automatically correct minor errors in a document while also prompting users with questions to deepen their thinking on the content.
While exploring extraheric AI designs that allow the coexistence with other AI types is beyond the scope of this paper, it presents a valuable opportunity for future research in human-AI interaction.

Extraheric AI is particularly applicable to domains where higher-order thinking skills determine the execution and outcome quality of a given task.
While Section~\ref{sec:interaction} illustrates applications demonstrated in the existing literature, examples of these applications would be: intellectual activities that involve opinion formation, decision making, and creative thinking (e.g., brainstorming, market analysis, and design proposals), resolving complex problems (e.g., programming and debugging a large project), and behavioral changes supported by perspective and perception changes (e.g., personal informatics for a person's physical and mental health).
For example, as Robins et al. note, \textit{``novice programmers must learn to develop models of the problem domain, the notional machine, and the desired program, and also develop tracking and debugging skills so as to model and correct their programs''}~\cite{robins2003learning}.
This combination of both conceptual and practical skills illustrates the essential nature of developing higher-order thinking skills for such complex task domains.
Future extraheric AI development should actively explore new application domains because higher-order thinking skills can be critical in a broad range of intellectual activities.

\section{Extraheric AI Interaction Strategies}
\label{sec:interaction}

While Section~\ref{sec:definition} clarifies the definition of extraheric AI, it is necessary to understand how researchers and developers can design and implement it.
As this paper emphasizes the HCI perspective of extraheric AI rather than machine learning models and architecture for it, this section discusses interaction strategies relevant to extraheric AI.
To explore these strategies in a comprehensive and bottom-up manner, we conducted an extensive literature survey, as outlined below.

We systematically collected full-paper publications at CHI 2023 and 2024 using the keywords: ``AI,'' ``artificial intelligence,'' ``LLM,'' and ``large language models'' in the ACM Digital Library.
We manually performed the initial screening based on the following criteria.

\begin{itemize}
    \item The paper demonstrates a prototype system utilizing AI technology (not limited to generative AI) or emulating AI (e.g., the wizard-of-Oz method), instead of executing purely observational studies. This criterion was essential as we aimed to focus on actual demonstrations of extraheric AI.
    \item The paper targets cognitive activities involving users' higher-order thinking. This criterion was chosen because such activities are the main focus of extraheric AI. 
    \item The paper demonstrates extraheric characteristics in its prototype design (i.e., promoting users' critical thinking). This criterion was included because it is the essential component of extraheric AI. The work rather categorized as pure supporting AI systems (i.e., orthotics, prosthetics, and exoskeleton) was excluded accordingly.
\end{itemize}

We then conducted a detailed review of the papers that satisfied all the three criteria to determine interaction strategies their prototypes as well as the application domains. 
To ensure objectivity and reliability, two of the authors independently reviewed these papers and annotated the interaction strategies they employed.
In cases of disagreement, we re-reviewed these papers to reach a consensus.
We also consolidated annotations that were closely related to enhance clarity and accuracy.

Through this process, we categorized 50 papers and identified eight distinct interaction strategies, which are discussed in detail below.
Table~\ref{table:interaction_strategy} illustrates our categorization against the interaction strategies and application domains.

\subsection{Suggesting \& Recommending}

Suggesting and recommending is an extraheric AI interaction strategy that involves proposing ideas, viewpoints, solutions, or actions to the user, without necessarily detailing the rationales behind them.
With this strategy, users' cognitive engagement comes in the form of evaluating and deciding whether or not to incorporate the AI's suggestions or recommendations into their thinking or tasks.
For example, in the context of news article reading, extraheric AI may recommend related articles with similar or different perspectives to encourage the user to explore multiple viewpoints.
In the context of technical tasks like software development, extraheric AI may suggest multiple implementations of a particular method and allow the user to choose the one they determine to be most appropriate.
An example system using this strategy is CoArgue, developed by Liu et al.~\cite{10.1145/3544548.3580932}, where the user can overview claims from different stances to construct their own arguments. 
In all such cases, it is critical that the AI makes \textit{multiple} suggestions to allow the user to evaluate and choose among them.

\begin{table*}[t]
    \small
    \centering
    \begin{tabular}{p{2.7cm}||p{1.15cm}|p{1.15cm}|p{1.15cm}|p{1.15cm}|p{1.15cm}|p{1.15cm}|p{1.15cm}|p{1.15cm}}
        \Hline
        & \rotatebox{90}{\textbf{Suggesting \&}} \rotatebox{90}{\textbf{recommending}} & \rotatebox{90}{\textbf{Explaining}} & \rotatebox{90}{\textbf{Nudging}} & \rotatebox{90}{\textbf{Debating \&}} \rotatebox{90}{\textbf{Discussing}} & \rotatebox{90}{\textbf{Questioning}} & \rotatebox{90}{\textbf{Scaffolding}} & \rotatebox{90}{\textbf{Simulating}} & \rotatebox{90}{\textbf{Demonstrating }} \\
        \hline \hline
        \textbf{Information seeking, information search} & \cite{10.1145/3613904.3642196}, \cite{10.1145/3613904.3642459}  & \cite{10.1145/3613904.3642059}, \cite{10.1145/3544548.3581369} & & & & \cite{10.1145/3544548.3581369} & & \\
        \hline
        \textbf{Dis/mis-information checking} & & \cite{10.1145/3613904.3641904}, \cite{10.1145/3544548.3581219}, \cite{10.1145/3613904.3642428},
        \cite{10.1145/3613904.3642805}  & & \cite{10.1145/3613904.3642513}, \cite{10.1145/3613904.3642545}  & & & & \\
        \hline
        \textbf{Textual content creation (e.g., writing)} & \cite{10.1145/3613904.3642867}, \cite{10.1145/3613904.3642134}, \cite{10.1145/3613904.3642549},  \cite{10.1145/3544548.3580957}, \cite{10.1145/3544548.3580907} & & & & & \cite{10.1145/3613904.3642134}, \cite{10.1145/3544548.3581225}, \cite{10.1145/3613904.3641899}  & & \\
        \hline
        \textbf{Discussions, argument construction} & \cite{10.1145/3613904.3642650}, \cite{10.1145/3544548.3580932}  & & & \cite{10.1145/3613904.3642322} & \cite{10.1145/3544548.3580672} & & & \\
        \hline
        \textbf{Personal informatics, behavioral changes} &  & \cite{10.1145/3613904.3642032} &  \cite{10.1145/3613904.3642790} & & & \cite{10.1145/3613904.3642081} & & \\
        \hline
        \textbf{Programming} & \cite{10.1145/3613904.3642229}, \cite{10.1145/3613904.3642773}  & \cite{10.1145/3544548.3580981}, \cite{10.1145/3613904.3642773}, \cite{10.1145/3613904.3642239}  & & & \cite{10.1145/3613904.3642349} & \cite{10.1145/3544548.3580981} & & \\
        \hline
        \textbf{Decision-making, sense-making, data analysis} & \cite{10.1145/3613904.3642480},  \cite{10.1145/3613904.3641891}, \cite{10.1145/3613904.3642024}, \cite{10.1145/3544548.3581469} &  \cite{10.1145/3613904.3642149}, \cite{10.1145/3613904.3642024},  \cite{10.1145/3613904.3641960}, \cite{10.1145/3544548.3581252}, \cite{10.1145/3613904.3642013}, \cite{10.1145/3544548.3581260} & & & & & & \\
        \hline
        \textbf{Education} & \cite{10.1145/3613904.3642647} & \cite{10.1145/3613904.3642887} & & & \cite{10.1145/3613904.3642887}, \cite{10.1145/3613904.3642041}, \cite{10.1145/3613904.3642806}, \cite{10.1145/3613904.3642647} & & \cite{10.1145/3544548.3581159} & \cite{10.1145/3613904.3642947}\\
        \hline
        \textbf{Design/idea/prototype explorations} & \cite{10.1145/3613904.3642794}, \cite{10.1145/3613904.3642908}, \cite{10.1145/3613904.3642216}, \cite{10.1145/3613904.3642698}, \cite{10.1145/3613904.3642901} & \cite{10.1145/3613904.3642216}, \cite{10.1145/3613904.3642335} & & \cite{10.1145/3613904.3642850} & & & & \\
        \Hline
    \end{tabular}
    \caption{Categorization of existing literature related to extraheric AI published as full papers at CHI 2023 and 2024 by interaction strategy and application domain. Note that some papers demonstrate multiple interaction strategies.}
    \label{table:interaction_strategy}
\end{table*}

\subsection{Explaining}

Explaining is a strategy in which extraheric AI offers explanations of information related to the task which the user currently engages in.
Unlike suggestions or recommendations, this strategy emphasizes providing details on the `why' and `how' of a particular piece of information.
In the context of news article reading for opinion formation, extraheric AI with this strategy may visualize additional background or contextual explanations about a particular component of the article the user is currently reading.
In this manner, extraheric AI allows the user to confirm their understanding and situate the article correctly.
ClarifAI, developed by Zavolokina et al.~\cite{10.1145/3544548.3581369}, demonstrates such an explaining strategy to alert users to potential propaganda content.
In the context of software development, extraheric AI may provide explanations of methods or API calls to allow the user to better understand the code's function.
It is important that extraheric AI offers explanations that allow the user to deepen their understanding of the task at hand.
It thus should aim to provide additional context or information rather than step-by-step instructions that the user may blindly follow.

\subsection{Nudging}

A nudge is an approach to subtly encourage or influence behavior through indirect suggestions and reinforcements without preventing alternative choices~\cite{thaler2008nudge}.
Although the concept originates in the field of behavioral economics, HCI research has extensively explored its applications and confirmed effects on decision-making and behavioral change~\cite{schneider2018digital,bergram2022digital}.
Extraheric AI using this strategy may indirectly show information that can persuade the user in particular directions while still offering them freedom to choose.
In the context of news article reading, extraheric AI may visualize a conceptual space of relevant articles in a side view, nudging the user to explore other opinions.
In the context of software development, a system could populate a dynamic list of relevant packages or libraries that may be of use to the user, but without specifically recommending any of them.
Wu et al. use this strategy in designing interventions for problematic smartphone use~\cite{10.1145/3613904.3642790}, utilizing large language models for constructing personalized and dynamic persuasive messages with different persuasion strategies.
It is important that systems using this strategy make it easy for the user to find information that helps them explore different directions and perspectives, without presenting their options as being necessary or exhaustive.

\subsection{Debating \& Discussing}

In this mode, users debate or discuss a given topic and exchange their thoughts and opinions with AI agents.
In the context of news article reading, extraheric AI using this strategy could offer an online discussion thread where the user may discuss their thoughts with AI agents holding various different opinions.
In the context of software development, the user may engage in paired programming with an AI peer, and discuss the use of different libraries, code structures, or algorithms.
Zhang et al. develop a system where the user can discuss with multiple AI agents exhibiting different viewpoints in order to deepen their understanding of different opinions and overcome filter bubble effects~\cite{10.1145/3613904.3642545}.
When using this strategy, it is important that debates and discussions focus on presenting different perspectives and ideas rather than simply disagreeing with or asking the user to justify their opinion.

\subsection{Questioning}

In this mode, extraheric AI asks questions about particular parts of what the user currently engages in.
Such questioning is not supposed to validate the correctness of opinions and perspectives, but rather stimulate users' cognitive activities to expand their thoughts or consider different perspectives.
In the context of news article reading, extraheric AI using this strategy may ask questions about a particular portion of the content, such as ``How do you think people in other countries perceive this news? What consequence could occur in their countries?''.
In the context of software development, extraheric AI may ask the user to explain how a particular code block functions, or why they chose to implement an algorithm in the way they did.
Danry et al. demonstrate the positive effects of this strategy on the critical thinking task of evaluating the logical soundness of statements that can create social division~\cite{10.1145/3544548.3580672}.
This strategy encompasses interactive environments based on a well-known effective pedagogical approach called \textit{learning by teaching}~\cite{duran2017learning}. 
Through users' active engagement with questions from AI agents, this process effectively stimulates users' higher-order thinking.

\subsection{Scaffolding}

Scaffolding is a learning approach where teachers offer temporary customized support to help students learn new concepts and skills, and gradually remove this help as students become more capable on their own~\cite{wood1976role}.
Scaffolding allows a learner to effectively complete tasks beyond their current ability level, but also, as Wood et al. note, results in \textit{``development of task competence by the learner at a pace that would far outstrip [their] unassisted efforts''}~\cite{wood1976role}.
Extraheric AI can serve as a scaffold for users by taking on part of a task and allowing them to focus only on particular portions at a time.
In the context of software development, extraheric AI using this strategy may help the user focus on program structure by allowing them to write pseudo-code or use visual programming methods before later translating these into functional code.
Lee et al. demonstrate DAPIE~\cite{10.1145/3544548.3581369}, where the AI agent offers step-by-step explanations while encouraging children to actively interact with it and assess their understanding.
However, as Wood et al. note, \textit{``comprehension of the solution must precede production. That is to say, the learner must be able to recognize a solution to a particular class of problems before [they are themselves] able to produce the steps leading to it without assistance}~\cite{wood1976role}.
As a result, it is critical that extraheric AI using this strategy focuses on developing the user's fundamental understanding of a task rather than simply allowing them to offload task decomposition.

\subsection{Simulating}

In this mode, extraheric AI simulates a circumstance where the user experiences a situation from a standpoint other than their own or develops skills that would be difficult to otherwise practice.
For example, AI agents could simulate audience members of different opinions and perspectives, allowing users to practice public speaking and responding to audience questions.
Extraheric AI could be tuned to different levels of aggressiveness to develop resilience and abilities for handling different types of audiences.
Simulations can also allow users to experience situations from a different standpoint.
For example, the user could take the role of an interviewer tasked with interviewing an AI agent playing the role of a job candidate.
By asking a variety of questions and observing the agent's responses, the user can think about how they may answer such questions as interviewees in actual job interviews.
The user will also have the opportunity to empathize with their interviewers, potentially giving them valuable insight into how to best communicate their ideas.
A series of art exhibitions by Lee et al. allowed students to immersively experience facets of AI (e.g., artwork generated using their facial data), stimulating critical thinking through an informal learning experience~\cite{10.1145/3544548.3581159}. 
As with other strategies, it is important that such simulations be designed to present a variety of viewpoints to encourage users to consider different perspectives and think critically about their own positionality.
This strategy could be particularly valuable for helping users understand their own and others' implicit biases.

\subsection{Demonstrating}

Demonstrating is a strategy where users simply observe the behavior or interaction of AI agents and learn implicitly through these observations.
In this case, there is no direct information flow from extraheric AI to users.
Users thus would have the largest freedom in how they interpret the behavior or interaction of AI agents and internalize take-aways through vicarious learning~\cite{roberts2010vicarious}.
In the context of news article reading for opinion formation, extraheric AI using this strategy may take the role of a peer, demonstrating their reading process and sharing opinions.
The user can review these demonstrations and construct their own opinions by integrating what they have observed with their own reading.
Liu et al. create a classmate AI agent in a virtual reality classroom that plays the role of an active student~\cite{10.1145/3613904.3642947}.
Students in the same virtual classroom observe its behavior, which can stimulate their active class engagement.
In addition to employing such a role model, extraheric AI using this strategy may include multiple AI agents to offer the demonstrations of diverse perspective or approaches to a task or topic.

\section{Extraheric AI Evaluation Approaches}
\label{sec:evaluation}

\begin{table*}[t]
    \small
    \centering
    \begin{tabular}{c|c|c}
        \Hline
        \textbf{Evaluation Area} & \textbf{Potential Evaluation Method} & \textbf{Reference} \\
        \hline \hline
        Germane Load & Revised \textit{NASA-TLX} & \cite{Gerjets2006_WorkedExamples} \\
        \hline
        Cognitive Activity (lower level) & Knowledge and comprehension tests & \cite{dorodchi2017wish, Jones2009} \\
        Cognitive Activity (inter. level) & Reflections; Concept maps & \cite{Dunfee2008, Watson2015} \\
        Cognitive Activity (higher level) & Performance-based assessment; Prototyping & \cite{abosalem2016assessment} \\
        \hline
        Sense of Agency & \textit{Sense of Agency Scale} & \cite{Tapal2017} \\
        Self-Efficacy & \textit{New General Self-Efficacy Scale} & \cite{chen2001validation} \\
        Task Motivation & \textit{Motivation Source Inventory} & \cite{Barbuto1998} \\
        AI Use Motivation & Measures of likeability and trust of agents & \cite{rau2009effects} \\
        Attribution of Credit and Blame & Measures of credit and blame & \cite{kim2006should} \\ \Hline
    \end{tabular}
    \caption{Areas of evaluation for extraheric AI and their potential metrics.}
    \label{tab:evaluation_approaches}
\end{table*}

User evaluation is a critical component in HCI research.
Qualitative evaluation approaches, such as journaling, questionnaires using open-ended questions, and interviews, can be widely applicable to extraheric AI research to examine its user experience and effects.
In this section, we present aspects of user experience and abilities that may convey the effects of extraheric AI, and discuss other possible approaches associated with germane cognitive load and Bloom's taxonomy~\cite{bloom1956taxonomy, anderson2001taxonomy}.
We, however, note that extraheric AI research can be highly dependent on its use contexts and applications, and thus the objectives of extraheric AI may be diverse.
As such, this section intends to lay a foundation to help researchers and developers explore how they can combine these approaches, as well as qualitative methods, to obtain a holistic perspective on user experience and the effects of their extraheric AI systems. 
We encourage the HCI research community to explore evaluation approaches together to acquire deep understanding of how extraheric AI influences users. Table~\ref{tab:evaluation_approaches} summarizes our proposed areas of evaluation and related metrics.

\subsection{Evaluations on Germane Load}

As discussed in Section~\ref{sec:definition}, extraheric AI can be interpreted as a mechanism to increase users' germane load.
It is thus important to assess how users' germane load would change with and without the presence of extraheric AI for completing a given task.
Although cognitive theory and educational psychology research have explored different approaches to evaluate the three types of cognitive load~\cite{DebueGermane2014, Orru2019CognitiveLoad}, HCI research has not adopted a standardized way of measuring cognitive load yet.
NASA-TLX~\cite{Hart1988_NASA_TLX} is a well-established metric for evaluating subjective workload and commonly used in HCI, but it is not considered associated with cognitive load theory.
To address this issue, Gerjets et al. revised NASA-TLX by introducing \textit{task demands}, \textit{navigational demands}, and \textit{efforts}, which are explicitly connected with intrinsic, extraneous, and germane cognitive load, respectively~\cite{Gerjets2006_WorkedExamples}.
These statements are: \textit{``how much mental and physical activity was required to accomplish the learning task (e.g., thinking, deciding, calculating, remembering, looking, searching etc.)''} for task demands, \textit{``how much effort the participant had to invest to navigate the learning environment''} for navigational demands, and \textit{``how hard the participant had to work to understand the contents of the learning environment''} for efforts.
As the NASA-TLX has already been widely adopted in the field of HCI, we suggest that this modified version may be immediately adapted to HCI research to quantitatively measure users' perceived cognitive load with extraheric AI.
However, in addition to the revised Gerjets et al.'s revised NASA-TLX, there exist other scales for measuring cognitive workloads~\cite{Orru2019CognitiveLoad}.
We encourage the HCI research community to discuss and explore how HCI research can adapt these scales to better assess users' germane load.

\subsection{Evaluations on Higher-order Thinking Skills}

Several theoretical frameworks exist in the fields of educational psychology and cognitive psychology to assess individuals' higher-order thinking skills.
For instance, Bloom's taxonomy~\cite{bloom1956taxonomy}, the SOLO Taxonomy~\cite{biggs1982solo}, Fink’s Taxonomy of Significant Learning~\cite{fink2003creating}, and the Paul-Elder Critical Thinking Framework~\cite{paul2013critical} can be potential frameworks for evaluating higher-order thinking skills. 
Here, we apply Bloom's taxonomy as an example to guide us on how to evaluate individuals' higher-order thinking skills with extraheric AI.

Bloom's taxonomy is a well-recognized framework that hierarchically organizes the stages of cognitive activities, and has already been widely adopted by HCI and computer science education research~\cite{Masapanta2018}.
The initial taxonomy was proposed by Bloom et al. in 1958~\cite{bloom1956taxonomy}, and includes six stages of cognitive development: \textit{knowledge} (remembering facts, terms, and basic concepts); \textit{comprehension} (understanding information); \textit{application} (using knowledge in new situations); \textit{analysis} (breaking information into components to understand its structure); \textit{synthesis} (combining different information and knowledge together to form a new solution); and
\textit{evaluation} (making judgments about information and ideas).
Anderson and Krathwohl revised it to emphasize creativity, application, and higher-order thinking skills~\cite{anderson2001taxonomy}: \textit{remember} (equivalent to \textit{knowledge} in the original Bloom's taxonomy), \textit{understand} (\textit{comprehension}), \textit{apply} (\textit{application}), \textit{analyze} (\textit{analysis}), \textit{evaluate} (\textit{evaluation}), and \textit{create} (\textit{synthesis}; it is placed as the highest level in the revised taxonomy).
While the order of the stages is slightly different, both taxonomies cover the same set of cognitive activities around higher-order thinking skills.
In this section, we discuss possible evaluation approaches following the categorization of the six stages of these taxonomies into three consolidated levels, following Jones et al.'s categorization~\cite{Jones2009}.

\subsubsection{Assessing Users' Knowledge and Comprehension Ability (Lower Level)}

\textit{Knowledge and comprehension} lie in the most basic levels in Bloom's taxonomy and its revised version.
They involve correct recall of facts, terms, and basic concepts, and the understanding of information by interpreting, summarizing, explaining, or translating it into one's own words.
Typical methods to evaluate individuals' understanding and knowledge involve assessing their recall of facts, terms, and basic concepts, as well as subject matter tests utilizing true or false questions, multiple-choice questions, fill-in-the-blank questions, summarizing, paraphrasing, and providing examples.
For instance, Jones et al. employed knowledge tests with questions phrased with ``define,'' ``list,'' ``state,'' ``identify,'' and ``label'' to gauge students' knowledge capability~\cite{Jones2009}.
Rubrics developed based on Bloom's taxonomy are another common approach for computer science educators to assess students' cognitive development~\cite{dorodchi2017wish}.

We note that extraheric AI research should not overemphasize supporting or assessing participants' lower-level cognitive abilities.
Knowledge is a fundamental tool for higher-order thinking but is not generally recognized as a higher-order thinking skill.   
Over-weighing changes in \textit{knowledge and comprehension} abilities thus may not capture the true effects and benefits of extraheric AI.

\subsubsection{Assessing Users' Application and Analysis Ability (Intermediate Level)}
\textit{Application and analysis} involve using knowledge and concepts in new situations, solving problems by implementing learned procedures or techniques, and being able to break down information into smaller components to understand its structure, relationships, and patterns.

\textit{Application and analysis} abilities can be assessed using a number of methods, including evaluating users' performance in a problem-solving context, and analyzing their reflections through reflective journals and workshops~\cite{Dunfee2008,Plack2007}.
Accuracy of concept maps, a visual tool that organizes and represents relationships between concepts, can also indicate the degree of individuals' \textit{application and analysis} abilities~\cite{novak1984learning}.
There exist various different approaches to evaluate concept maps~\cite{Watson2015, McClure1999}, and choice of these evaluation methods may depend on research objectives and contexts.
Future researcher is encouraged to explore how extraheric AI may utilize these and other methods.

\subsubsection{Assessing Users' Evaluation and Creation Ability (Higher Level)}
\textit{Evaluation and creation} involve making judgments or decisions based on criteria and standards by assessing the value or quality of ideas, methods, or materials, as well as combining elements in new ways to form a coherent outcome, generating novel ideas, or creating original products.

Performance-based assessment, such as analyzing the process and outcome of users' discussion, debate, and decision-making tasks is appropriate for assessing their higher-level abilities~\cite{abosalem2016assessment}.
By analyzing arguments provided in a discussion or written report, researchers can gauge users' higher-level cognitive abilities with regard to identifying the strengths and weaknesses of arguments and counterarguments based on evidence and logic.
Inviting users to outline alternative solutions or methods to complete the given task is also effective in assessing their \textit{evaluation and creation} abilities.
We suggest that the HCI research community, which has a long history of engaging participants through interactive and participatory interface prototyping approaches, can uniquely contribute to assessing \textit{evaluation and creation} abilities by exploiting these existing approaches and their associated body of knowledge.

\subsection{Attitudinal and Behavioral Metrics}
\label{sec:eval_metrics}

In addition to the direct examination of higher-order thinking skills discussed in the previous section, changes in users' attitudes and behavior over time can also be an important indicator of the effects of extraheric AI.
For instance, users may exhibit a more open attitude toward diverse opinions or an increase in self-efficacy after they have developed stronger higher-order thinking skills.
There exist established scales for quantitatively examining these changes.
We point out key attitudinal and behavioral aspects that are relevant to the use of extraheric AI, and invite HCI researchers to broaden the exploration of methods to assess them.

\subsubsection{Sense of Agency}

Sense of agency refers to the feeling or perception of having control over one's own actions, thoughts, and effects.
We expect that the development of higher-order thinking skills may lead users to develop a stronger sense of control over their tasks and the information presented to them.
Such perception would accordingly be reflected in their sense of agency.
Accordingly, Xiao et al. included a sense of agency measurement in their extraheric AI research to support users to achieve better-informed consent~\cite{10.1145/3544548.3581252}.
Beyond the effects of higher-order thinking skill development, we also hypothesize that extraheric AI may have different impacts on users' sense of agency compared to orthotics, prosthetics, and exoskeleton.
Systems that negatively impact human agency appear to have a higher chance of contributing to feelings of dehumanization~\cite{KnowledgeWorkerCHI2024}, reduced meaningful human interaction~\cite{wagner2019liable}, and reduced trust in AI~\cite{liu2021ai}.
This suggests that measuring and understanding users' sense of agency is critical for the success of future human-AI interaction paradigms.

Although measurement of agency remains a challenging task with considerable disagreement with regard to its methodology~\cite{cavazzoni2022we}, there exist several well-established scales.
The Sense of Agency Scale developed by Tapal et al.~\cite{Tapal2017} is one of the most commonly-used instruments in HCI and human-AI interaction research~\cite{DraxlerTOCHI, 10.1145/3544548.3581252}.
Exploring appropriate adaption of existing sense of agency scales to extraheric AI research is an important open research direction.

\subsubsection{Self-efficacy}

Self-efficacy refers to one's perception of the ability to effectively utilize tools and environments to achieve desired outcomes~\cite{Bandura1977}.
While sense of agency is about the perception of control over given systems and information, self-efficacy is about confidence in the ability to complete tasks.
In extraheric AI research, the development of higher-order thinking skills may contribute to a deeper understanding of the process of completing tasks, potentially resulting in improved self-efficacy.
As noted above, one of the major concerns with over-reliance on AI is the risk of deskilling and reduced cognitive engagement.
These concerns relate directly to users' self-efficacy, and assessing changes in self-efficacy is therefore critical for understanding the effectiveness of extraheric AI systems.
Similarly to sense of agency, there exist several scales for self-efficacy.
The New General Self-Efficacy Scale developed and further validated by Chen et al. is one of the established scales~\cite{chen2001validation}, though it has not yet been widely adopted by HCI research.
The HCI community is encouraged to further explore how existing self-efficacy scales can be adapted to extraheric AI research.

\subsubsection{User Motivations and Willingness}

As extraheric AI may increase germane cognitive load, users may experience a cognitive burden, particularly when first using a system.
While this may potentially decrease their motivation, extraheric AI may also strengthen their motivation when they feel the sense of growth from the cultivation of their higher-order thinking skills.
Understanding how users' motivational changes over time can help researchers better adapt the design of extraheric AI systems. Several theories and measurements have been widely used in HCI research to investigate users' motivation for using new technology. 
These include Self-Determination Theory (SDT)~\cite{deci2000and}, the Technology Acceptance Model (TAM)~\cite{davis1989technology, hornbaek2017technology}, the Motivation Source Inventory~\cite{Barbuto1998}, or the Intrinsic Motivation Inventory~\cite{mcauley1989psychometric}. 
As the fluctuation of users' motivation when working with extraheric AI remains unexplored, we encourage researchers to consider this aspect when assessing the effectiveness of extraheric AI.

Another important evaluation metric related to motivation is user willingness to use extraheric AI.
As extraheric AI may not offer direct support for a given task, users may exhibit a lower willingness to continue to use. 
However, a decrease in users' willingness may also occur when they have become proficient in the task without the use of AI support, which can indicate a move towards positive disengagement.
Understanding how user willingness changes alongside developments in their higher-order thinking skills is critical for understanding the effects on extraheric AI on users' attitudinal and behavioral changes.

\subsubsection{Responsibility and Ownership Attribution with Extraheric AI}

The magnitude of effort and autonomy users perceive in interaction with extraheric AI may influence their perceived attribution of responsibility and ownership of final outcomes to AI agents.
Past studies have shown that users attributed more blame and responsibility when encountering failed collaborative outcomes to social actors that they conceptualized as more intelligent~\cite{awad2020drivers, miyake2019mind}.
Accordingly, Kadoma et al. used the perceived ownership between users and AI in co-writing process as one of the evaluation metrics~\cite{10.1145/3613904.3642650}.
As extraheric AI aims at stimulating users' high-order thinking skills during task completion, it is essential to know how users assign responsibility or credit for the collaborative outcome to their extraheric AI agents, especially when both a successful and failed outcome can occur.
A common approach to assess responsibility attribution is to ask users to indicate the amount of blame and credit each stakeholder should receive for a specific task~\cite{kim2006should, feather1969attribution}.
A question asking the extent to which users feel that the outcome is attributed to them can also measure perceived ownership~\cite{10.1145/3613904.3642650, DraxlerTOCHI}
As users may engage in many activities that require higher-order thinking skills, understanding how users attribute the collective responsibility and credit over time is critical for extraheric AI research.

\section{Extraheric AI Design Considerations}

Existing literature has identified and validated guidelines and design considerations for human-AI interaction research~\cite{SaleemaCHI2019, TianyiCHI2023, NurCHI2023}.
While these guidelines and design considerations also apply to extraheric AI, its unique characteristics introduce additional design considerations researchers and developers should take into account.

\subsection{Explicitly Considering The Social Roles of Extraheric AI Agents}
\label{sec:social_roles}

As users are expected to collaboratively explore various perspectives and information with extraheric AI, AI agents may play different social roles similar to what is seen in human-human communication and collaboration.
K{\"o}bis et al. described four social roles of AI~\cite{Kobis2021BadMachines}: role model, advisor, partner, and delegatee\footnote{Their original article~\cite{Kobis2021BadMachines} used the term of ``delegate'' to represent AI use where people outsource their tasks to AI. We decided to use ``delegatee'' to clarify the relationship between users and AI.}, and discuss how different roles may cause different possible risks of negatively influencing users' ethical behavior.
With extraheric AI, it is crucial to explicitly consider agents' expected or perceived social roles, as users may interpret the same output differently depending on the roles these AI agents have.
There may also be other possible social roles in the context of extraheric AI; for example, a competitor role where an AI agent competes against users may contribute to users' active thinking.
Kim et al. examined the effects of a social bot playing a role of a depressed peer, which displays depressive symptoms to urge users to offer support and encouragement~\cite{Kim2020Helping}.
The interactions with the bot helped their study participants reframe their own negative experiences.
Future extraheric AI research should consider both positive and negative effects amplified by these social roles on the effectiveness and acceptance of the developed systems more broadly.

Uncovering the design space of extraheric AI social roles is therefore an important research agenda.
The effect of such social roles on user outcomes needs further examination through empirical studies.
For example, students may benefit more from interacting with extraheric AI agents playing the role of their peers rather than teachers, as Liu et al. demonstrated that an AI agent that simulates an active student peer in a virtual classroom can promote students' class participation~\cite{10.1145/3613904.3642947}.
Future research on extraheric AI should consider not only the interaction strategies by which AI can promote higher-order thinking skills, but also how the social roles of AI agents can enhance or degrade this process.

\subsection{Generating Diversified Outputs}
\label{sec:diversified_outputs}

To promote higher-order thinking skill development, the output of extraheric AI should be designed to encourage users' diverse interpretations instead of constraining users to particular directions or perspectives.
In particular, suggestions from AI have already been found to produce idea fixation and discourage divergent thinking~\cite{wadinambiarachchi2024effects}.
Presenting multiple diversified outputs is thus critical, particularly when extraheric AI employs the \textit{suggesting \& recommending} interaction strategy.
This design consideration is also crucial for the \textit{debating \& discussing} interaction strategy.
Only employing an extraheric AI agent with a very similar stance to users may create undesirable echo-chamber effects, whereas multiple AI agents with diverse opinions have already been shown to help mitigate filter bubble effects~\cite{10.1145/3613904.3642545}.
It is essential that future extraheric AI research explores how a system can produce diversified responses and perspectives, and present them in ways that push users to develop and use their higher-order thinking skills.
However, when designing for diversity, it is also critical that designers and developers be aware of their positionality and potential blindspots~\cite{scheuerman2024products}, as well as continue to stay abreast of challenges with AI bias and other ethical issues~\cite{jiao2024navigating}.

\subsection{Maintaining Non-judgmental Attitudes and Behavior}

Extraheric AI should encourage users' intellectual exploration.
And in many cases, there are no clear right or wrong directions for users to explore.
It is thus critical that extraheric AI systems remain non-judgmental, and in particular, avoid providing simplified evaluations (e.g., numerical scores) of users' outcomes.
Non-judgmental behavior is a fundamental principle of motivational interviewing, a communication approach designed to facilitate and engage intrinsic motivation within users to promote positive behavioral change~\cite{miller2012motivational}.
Extraheric AI should embrace how users integrate different perspectives into their tasks, including their decisions to not make use of AI output.
Non-judgmental behavior thus can be beneficial for promoting honesty and trust toward extraheric AI.  

Researchers and developers may want to perform some quantification of users' outcomes to evaluate their extraheric AI systems.
However, we maintain that such quantification should only be used for evaluating system performance (e.g., effects of different interaction strategies or roles on users' higher-order thinking skills), rather than for scoring users or otherwise directly rating their `performance'.
In general, extraherics is not primarily about directly accelerating task completion, but rather about expanding users' cognitive capabilities.

\subsection{Aligning with Users' Workflows and Contexts for Task Completion}
Integration with existing workflows has been identified as  the key to adoption of a variety of technologies, ranging from creativity support tools~\cite{palani2022don} to generative AI~\cite{russo2024navigating}.
While extraheric AI aims at fostering users' higher-order thinking skills, it should also support the tasks users wish to complete.
It thus must take into account how it can provide users with opportunities to expand their higher-order thinking skills while minimizing interference with their task workflows.
This constitutes an important difference from learning systems including ITS, where the primary focus lies in skill training and development.
Understanding users' existing workflows and contexts is critical for designing appropriate extraheric AI that can maintain a balance between supporting users' task completion and cognitive development.
Formative studies can help to establish such understandings while iterative design explorations or co-design processes are effective approaches for ensuring that extraheric AI systems match users' workflows and contexts.

\subsection{Embracing User Disengagement from Extraheric AI}
As users engage with extraheric AI over time, they may develop sufficient higher-order thinking skills, and as Lewis and Smith note, may become sufficiently adept at the task at hand such as to no longer require the use of these skills to complete it~\cite{lewis1993defining}.
For example, in the context of writing, users may initially obtain help from extraheric AI to engage in critical thinking by being questioned about the content of a particular sentence or paragraph, but may become able to do so independently later.
In such cases, users may discontinue their use of extraheric AI as they feel stronger self-efficacy in the given task.
While adaptively changing the interaction strategies or social roles of extraheric AI can be a valid approach to encourage continued engagement, disengagement can also be considered indicative of sufficient higher-order thinking skill development.
As a result, future research should not overemphasize the frequency or occurrences of extraheric AI use in its evaluations.
Section~\ref{sec:evaluation} discusses evaluation approaches for extraheric AI, and advocates for a holistic perspective for measuring the different aspects of how people use extraheric AI systems.
While the frequency of extraheric AI use is one potential metric to understand user behavior, researchers and developers should carefully interpret and thoroughly discuss its contextual significance, instead of simply observing its magnitude. 
We therefore encourage applying mixed methods using diverse data sources to understand users' motivations for both engagement and disengagement.

\section{Research Opportunities}

The introduction of extraheric AI has the potential to contribute to and extend existing understanding and theory about human-AI interaction, and the use of extraheric AI may change the way people internalize knowledge and develop higher-order thinking skills. 
Here, we outline several open directions for studying extraheric AI, from interaction and interface design, and cognitive load theory, to new evaluation metrics, and interpersonal and social implications.

\subsection{Technology to Present Multiple Responses and Perspectives}
A first set of research questions concerns how to design technology that could present multiple responses and perspectives from extraheric AI.
In addition to explorations of interface designs that would not overwhelm users or interfere with their current tasks, diversifying responses and perspectives from extraheric AI is also an important technical research challenge.
Information retrieval research has introduced the concept of search result diversification~\cite{RakeshWSDM2009}, which refers to the process of delivering a set of relevant searches that also cover a broad range of aspects or perspectives related to the search query.
This concept is particularly important in scenarios where the search query may be ambiguous, or where different users may have different information needs.
While the goal of search result diversification lies in maximizing the likelihood that users can find what they are looking for, output diversification of extraheric AI aims at increasing the likelihood of users engaging in higher-order thinking activities. 
Unlike search results, which themselves are not directly controllable by search engines, researchers and developers have the freedom to control the output from extraheric AI agents by configuring their characters, personality, social roles, and behavior.
Employing multiple extraheric AI agents can be beneficial with respect to generating multiple perspectives and opinions, contributing to output diversification as a whole.
Future research is encouraged to explore creative approaches for extraheric AI output diversification by exploring various AI agent designs.

\subsection{HCI Perspectives on Germane Cognitive Load}

In Section~\ref{sec:definition}, we discussed the difference between extraheric AI and other supportive AI through the lens of the cognitive load theory.
Germane cognitive load is a relatively new concept, and remains controversial within the field of cognitive science.
Kalyuga, for example, critiques the three-way model of cognitive load and proposes that it is more of an aspect of intrinsic load~\cite{kalyuga2011cognitive}.
Similarly, de Jong points out the blurred nature of the boundaries between these types of cognitive load and discusses the challenging nature of separating them~\cite{dejong2010cognitive}.
We argue that the HCI research community can contribute to further understanding these types of cognitive load by employing actual interactive systems and performing evaluations (e.g., comparative studies between extraheric and non-extraheric AI systems).
However, we also echo de Jong's discussion on the challenges of evaluating solely one of these cognitive load types, and note that careful consideration is needed for future research.
We see an opportunity for the HCI research community to contribute to the discussion of cognitive load theory in the context of extraheric AI by incorporating holistic evaluation metrics that assess both human and system performance.
We believe that extraheric AI research can create a new bridge between computer science and cognitive science, calling for new interdisciplinary collaboration.

\subsection{Explorations of Alternative Evaluation Approaches for Extraheric AI}

In Section~\ref{sec:evaluation}, we discussed possible evaluation approaches based on cognitive load theory and Bloom's taxonomy, as well as attitudinal and behavioral scales.
As extraheric AI is a new concept, we have borrowed existing evaluation metrics from the field of education research.
However, we also suggest that there are unique opportunities for the HCI research community to develop alternative frameworks and evaluation approaches to capture changes in users' high-order thinking skills over time when using extraheric AI.

As the development of higher-order thinking skills may need long-term effort, future extraheric AI research will need to devise tracking methods for cognitive development and evaluate their validity.
Periodical evaluations employing approaches discussed in Section~\ref{sec:evaluation} may be a possible design, but researchers and developers should also consider the effort required of participants for repeatedly completing such evaluation tasks.
Observing the evolution of reflection journals or concept maps can be an interesting approach to gauging the development of higher-order thinking skills.
For example, researchers and developers may quantify the correctness and level of detail of concept maps developed by participants and observe how these metrics transition over time~\cite{doyle2022eliciting}.
The HCI research community has extensive experience in in-the-wild, long-term user evaluations, and we argue that this knowledge can contribute to establishing evaluation methods for longer-term development of higher-order thinking skills in a realistic setting.

\subsection{Effects Caused by Different Social Roles of Extraheric AI}
\label{sec:RO_social_roles}

As discussed in Section~\ref{sec:social_roles}, the social roles of extraheric AI agents may have impacts on user perception and behavior of these agents, and, accordingly, have effects on encouraging cognitive activities related to higher-order thinking skills.
As demonstrated through studies of interpersonal interaction, the perceived social roles of interlocutors can influence behavior~\cite{zimbardo1995psychology}, power dynamics~\cite{ely1995power}, and communication patterns~\cite{welser2007visualizing}.
However, it is still unclear how interacting with extraheric AI designed with varying social roles may influence these aspects and further users' higher-order thinking skills.
Examining the effects of different social roles of extraheric AI is therefore a key open research direction.

Multiple AI agents acting in different social roles have the potential to enable social learning for users as well as produce diversified outputs, as suggested in Section~\ref{sec:diversified_outputs}.
Social learning involves acquiring knowledge, skills, or behavior by observing, imitating, and interacting with others in a social context~\cite{bandura1977social}. 
Individuals may gain higher-order thinking skills through collaborative interactions, discussions, and the exchange of diverse perspectives with other people.
However, whether and how a similar social learning effect can be attained when users are interacting with multiple extraheric AI agents in a simulated social context is largely unknown. 
Therefore, future research efforts can investigate the effects caused by interacting with multiple extraheric AI agents that display different social roles.

\subsection{Ethical Considerations and Guidelines for Extraheric AI}

Recent years have seen substantial investigations and reflections on the ethical issues of machine learning~\cite{mittelstadt2016ethics}, large language models~\cite{jiao2024navigating}, and other forms of machine ethics~\cite{anderson2011machine}.
However, as Elliott et al. note, these numerous guidelines can be practically counterproductive and require harmonization to be usable~\cite{elliott2021towards}.
The development of extraheric AI and its application to various domains can and must consider such ethical issues and guidelines in its implementation, but, as a new proposal, it will likely introduce new ethical considerations as well.
For example, extraheric AI may introduce competing incentives for users and developers as it potentially leads to user mastery and thus disengagement from AI.
Many existing AI products rely on a business model of continuous use, potentially creating a trade-off between users' learning outcomes and providers' financial outcomes in the case of extraheric AI.
In the past, the development of many AI technologies has preceded the investigation of ethical concerns, which has been largely relegated to fields outside computer science. 
We urge researchers exploring extraheric AI to proactively integrate ethical considerations into their research programs from the start, in order to develop and integrate design and ethical best practices simultaneously.

A set of related open questions for extraheric AI design concerns user populations that may need extra caution and care.
For instance, in light of existing childhood development research, extra caution may be necessary in the application of extraheric AI to educational contexts involving children or adolescents.
Research on children's free play and its impacts on mental and physical development suggests that undirected play undertaken freely by children is essential for the development of physical and cognitive abilities and mental well-being~\cite{gray2011decline}. 
During group and collaborative activities with others, children develop numerous competencies beyond the task at hand, which allow them to \textit{``tolerate bruises, handle their emotions, read other children's emotions, take turns, resolve conflicts, and play fair''}~\cite{haidt2024anxious}.
However, these educationally-beneficial interactions may not occur effectively through interaction with AI agents, which can be easily ignored, shut off, or reset without any consequences, potentially counteracting the educational benefit that extraheric AI would otherwise provide.
Similarly, other specific user populations and application domains may require extra caution to design ethical extraheric AI, and future research is encouraged to explore domain-specific considerations and guidelines.

\section{Conclusion}

This paper presents \textit{extraheric} AI, a novel human-AI interaction conceptual framework aimed at mitigating the risks of over-reliance on AI, which can lead to human deskilling and reduced cognitive engagement.
Unlike traditional AI designs that replace or augment human cognition, extraheric AI fosters higher-order thinking skills by engaging users through questions and alternative perspectives rather than providing direct answers and support to given tasks.
This paper illustrates HCI research components of extraheric AI: interaction strategies, evaluation approaches, and design considerations.
As discussed above, extraheric AI opens up several research opportunities on which the HCI research community can take a strong initiative.
We hope that this work will serve as a catalyst for deeper discourse and further research on human-AI interaction that prioritizes a balanced partnership between humans and interactive intelligent systems.

\begin{acks}
We appreciate all the support we received in preparing this paper. 
In particular, Barnaby Ralph participated in our exploration of a new term for our conceptual framework and eventually coined \textit{extraherics}. He also contributed to our discussions of extraherics by bringing in posthuman literature.
Kakeru Miyazaki assisted us with conducting a literature survey to identify extraheric AI strategies and application domains. 
We greatly appreciate their dedicated help.
This project was partly supported by JST PRESTO (Grant Number: JPMJPR23IB).

\end{acks}

\bibliographystyle{ACM-Reference-Format}
\bibliography{references}

\appendix









\end{document}